\documentclass[twocolumn,aps,pre,nofootinbib,floatfix,preprintnumbers,amsmath,amssymb,superscriptaddress,10pt,colorlinks,a4paper,linenumbers]{revtex4-1}

\usepackage{ulem}
\renewcommand{\emph}{\textit}

\usepackage{selinput}

\usepackage[left=1.6cm,right=1.6cm,bottom=1.7cm,top=1.7cm]{geometry}

\usepackage[usenames,dvipsnames]{color}
\usepackage{natbib}
\usepackage{dsfont}
\usepackage{xr}
\usepackage{tensor}

\usepackage[T1]{fontenc}
\usepackage[english]{babel}

\usepackage{lmodern}
\usepackage{textcomp}		
\usepackage{textgreek}		

\usepackage{enumerate}

\usepackage{amsmath}		
\usepackage{amssymb}		
\usepackage{amsfonts}
\usepackage{mathrsfs}
\usepackage{braket}
\usepackage{mathtools}
\usepackage{siunitx}

\usepackage{bm}				

\usepackage[version=4]{mhchem}			

\usepackage{graphicx}	
\usepackage[export]{adjustbox}

\usepackage{lettrine}

\usepackage[english]{varioref}

\usepackage[urlcolor=RoyalBlue,citecolor=RoyalBlue,linkcolor=Black]{hyperref}

\usepackage{esvect}

\newcommand{\mean}[1]{\left < #1 \right >}

\renewcommand{\epsilon}{\varepsilon}
\renewcommand{\theta}{\vartheta}
\renewcommand{\phi}{\varphi}

\renewcommand{\vec}[1]{\mathbf{ #1 }}

\makeatletter
\renewcommand\section{%
  \@startsection
    {section}%
    {2}%
    {-10pt}%
    {-1.5em}%
    {3pt}%
    {\sffamily\large\bfseries}%
}%
\makeatother

\makeatletter
\renewcommand\paragraph{%
  \@startsection
    {paragraph}%
    {3}%
    {-\parindent}%
    {-1.8em}%
    {0.5pt}%
    {\sffamily\normalsize\bfseries}%
}%
\makeatother

\makeatletter
\newcommand\metparagraph{%
  \@startsection
    {paragraph}%
    {3}%
    {\parindent}%
    {0em}%
    {-0.6em}%
    {\newline \sffamily\small\bfseries}%
}%
\makeatother

\renewcommand{\figurename}{Figure}

\makeatletter
\renewcommand{\fnum@figure}{\small{\sffamily{\textbf{\figurename~\thefigure}}}\normalfont}
\renewcommand*{\@caption@fignum@sep}{ \small{$\boldsymbol{|}$} }
\makeatother

\newcommand{\fighead}[1]{\small{\sffamily{\textbf{#1}}}\normalfont}

\newcommand{\balancepage}[0]{\onecolumngrid\newpage\twocolumngrid}

\newcommand{\ts}{\textsuperscript}

\begin{document}
\title{Cargo size limits and forces of cell-driven microtransport}

\author{Setareh~Sharifi~Panah}
\affiliation{Institute of Physics and Astronomy, University of Potsdam, Karl-Liebknecht Stra{\ss}e 24/25, 14476 Potsdam, Germany}

\author{Robert Gro{\ss}mann}
\affiliation{Institute of Physics and Astronomy, University of Potsdam, Karl-Liebknecht Stra{\ss}e 24/25, 14476 Potsdam, Germany}

\author{Valentino Lepro}
\affiliation{Institute of Physics and Astronomy, University of Potsdam, Karl-Liebknecht Stra{\ss}e 24/25, 14476 Potsdam, Germany}

\author{Carsten Beta}
\email{beta@uni-potsdam.de}
\affiliation{Institute of Physics and Astronomy, University of Potsdam, Karl-Liebknecht Stra{\ss}e 24/25, 14476 Potsdam, Germany}


\begin{abstract}
The integration of motile cells into biohybrid microrobots offers unique properties such as sensitive responses to external stimuli, resilience, and intrinsic energy supply.
Here we study biohybrid microtransporters that are driven by amoeboid {\it Dictyostelium~discoideum} cells and explore how the speed of transport and the resulting viscous drag force scales with increasing radius of the spherical cargo particle.
Using a simplified geometrical model of the cell-cargo interaction, we extrapolate our findings towards larger cargo sizes that are not accessible with our experimental setup and predict a maximal cargo size beyond which active cell-driven transport will stall.
The active forces exerted by the cells to move a cargo show mechanoresponsive adaptation and increase dramatically when challenged by an external pulling force, a mechanism that may become relevant when navigating cargo through complex heterogeneous environments.
\end{abstract}
\nolinenumbers
\maketitle


\section*{Introduction}

Soft-bodied micromachines with bio-inspired modes of locomotion, such as crawling or swimming, are essential to fulfill many demanding mechanical tasks on the micron scale, including targeted drug delivery.
Examples include both synthetic~\cite{Katuri2016,Sitti2009} as well as biohybrid microcarriers, mostly based on cellular microswimmers~\cite{Akolpoglu2022,jin_collective_2021,Xu2017,Park2017,Sokolov2009,Ahmad2022}.
The ability to maneuver through confined, structured terrains~\cite{Doshi2011,Xu2017,Wu2022}, the multi modal locomotion on surfaces with varying adhesion properties~\cite{Lu2018,Hu2018}, and the targeted delivery of cargo particles~\cite{Nagel2018,Lepro2022} are examples of recent advancements in developing soft, crawling micromachines.
Yet, many challenges remain, including questions of power supply, sensing capacities, and long-term retention that are common to many small-scale robots~\cite{Lu2018,Sitti2009,Wang2013,Carlsen2014,Sun2020,Ceylan2019}.
Ideally, the unique energy efficiency, along with the integrated sensing machinery of biological cells, can be directly harnessed in a biohybrid approach, where motile cells are combined with synthetic components to a functional device~\cite{jin_collective_2021,Xu2017,Sun2020,Lee2022,Alapan2019,Ceylan2019}.

In this spirit, motivated by the motile capacities of amoeboid cells and by their widespread occurrence~\cite{Titus2017}, we recently proposed a biohybrid microcarrier that operates by directly loading a piece of microcargo onto a motile amoeboid cell~\cite{Nagel2018,Lepro2022}.
As the active driving element, we used cells of the social amoeba \textit{Dictyostelium discoideum} (\textit{D.~discoideum}) that carried different micronsized cargo particles.
Owing to the highly nonspecific adhesive properties of these cells~\cite{Loomis2012,Kamprad2018}, the physical link between the cargo and the carrier is established spontaneously upon collision, without additional surface functionalizations.
The cargo is then subjected to the forces exerted by the motile cell leading to displacements of the cargo.
As the crawling locomotion of {\it D.~discoideum} shares many similarities with the motility of leukocytes that travel through narrow, confined environments during an inflammatory response~\cite{Artemenko2014,Friedl2001,Titus2017,IshikawaAnkerhold2022,Shao2017,Xu2021}, it makes them a valuable model organism to study the transport capacities of motile eukaryotic cells.

In this work, we explore the potentials and limitations of this biohybrid transport system, which we will also refer to hereafter as ``cellular truck''.
We first investigate the active forces exerted by the cell on the cargo as well as the limiting cargo size for microparticle transport in an open, isotropic fluid environment.
Using high-speed live cell imaging, we show that only minimal forces up to average values of around~$0.4\, \mbox{pN}$ are exerted on spherical cargo particles.
Based on a simplified geometrical model for the cell-cargo interaction, we estimate that beyond a limiting cargo radius of about $113\, \mbox{\textmu m}$, cells will, on average, no longer displace the particle.
Finally, we use a microfluidic chamber to expose the cellular truck to a Poiseuille flow that allows us to probe the response of the truck to an external force pulling on the cargo particle.
Here, we measure significantly larger forces of up to 0.5~nN, suggesting that the forces generated by the cell may adapt and significantly increase when challenged by an external impact.

\begin{figure*}[t]
\centering
\includegraphics[width=.8\textwidth]{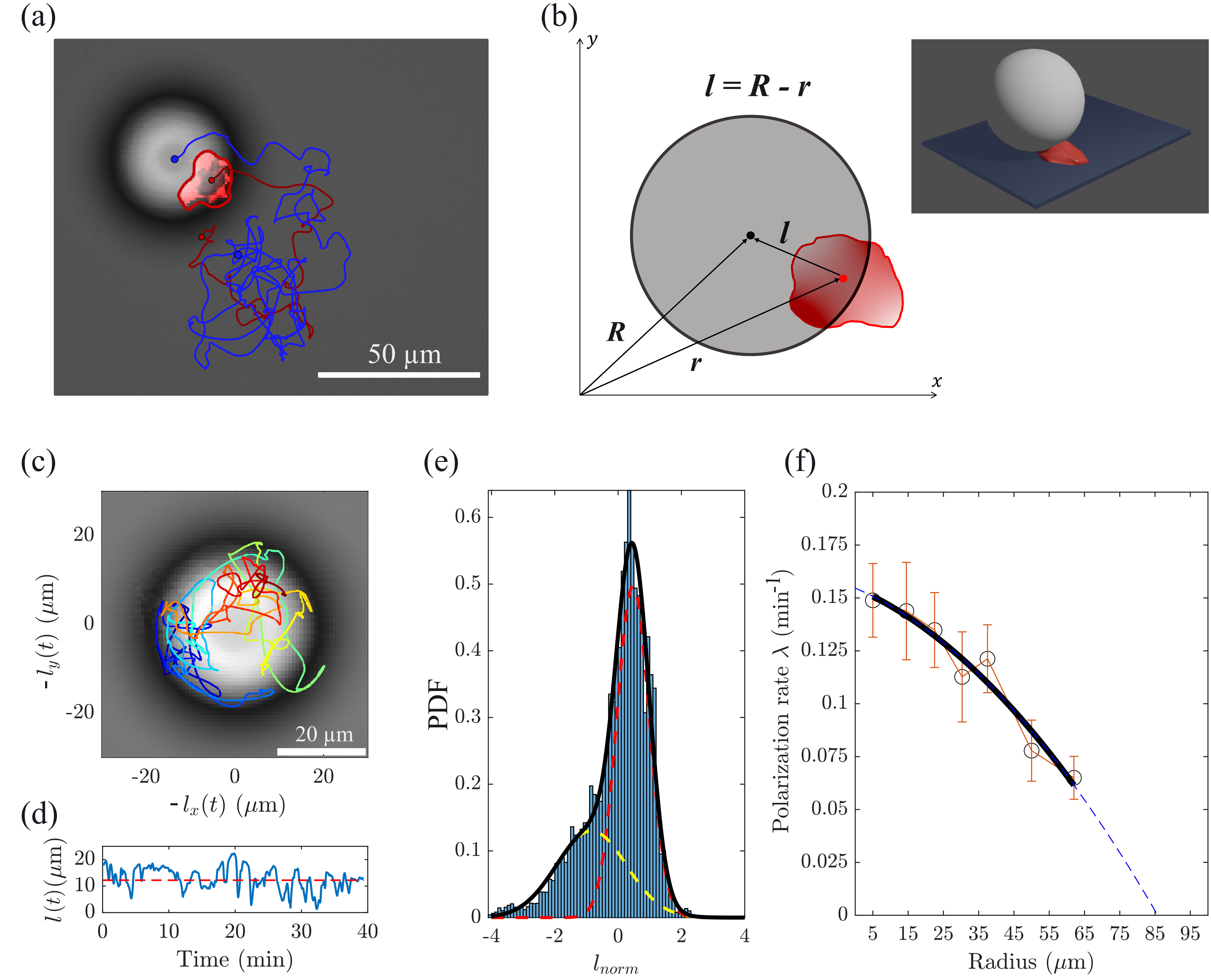}
\vspace{-0.2cm}
\caption{\fighead{Illustration and characteristic dynamics of the cellular truck.} (a)~An example of a microscopy image of the cellular truck along with trajectories of the cell and the cargo depicted in red and blue, respectively (Movie~2). The red fluorescence signal marks the F-actin density of the cell. The segmented cell contour is shown with a solid red line. The light gray sphere, encircled with a darker ring in the bright-field image, is the cargo particle with a radius of 23~\textmu m. (b)~Main geometrical quantities of the system. The vectors~$\boldsymbol{r}$ and~$\boldsymbol{R}$ denote the cell and the cargo positions, respectively. The vector~$\boldsymbol{l}=\boldsymbol{R}-\boldsymbol{r}$ quantifies the relative cargo position with respect to the cell. A three dimensional rendering of the cellular truck is provided as an inset. (c)~Dynamics of the cell in the frame of reference of the cargo, with the microscopy image of the cargo particle in the background. The transitions can be distinguished from episodes of circling motion. The time is color-coded from blue (beginning) to red (end). (d)~Temporal changes in the distance~$l(t) = \left | \boldsymbol{l} (t)\right |$ between the cell and the cargo centers of mass, shown for a particle with a radius of 23~\textmu m. The red dashed line marks the equilibrium distance~$l_{0}$. The polarization rates shown in (f) are derived from the frequency of the peaks extending below~$l_{0}$. (e)~Histogram of the normalized cell-cargo distances~$l_{norm} = \left ( l - \mean{l}\right )\!/\mbox{std}(l)$ for cellular trucks with a cargo of radius~$23~\mbox{\textmu m}$. In order to correct for cell-to-cell variability, the average cell-cargo distance~$\mean{l}$ was subtracted from the recorded time series~$l(t)$ and, subsequently, divided by the standard deviation for each trajectory. A double Gaussian distribution is fitted to the asymmetric shape of the histogram (black solid line), capturing the two states of the system dynamics. The major fraction of~$l$ values belongs to the idling rest state with values fluctuating around~$l_{0}$ (peak of the dashed red Gaussian); a second fraction, covering smaller~$l$ values, represents the transitions states (dashed yellow Gaussian). (f)~Monotonous decay of the polarization rate~$\lambda$ as a function of particle radius along with a parabolic fit (solid black line). The blue dashed line is the extension of the parabolic fit up to a particle radius of around 85~\textmu m, where the transition rate decays to zero. See Ref.~\cite{Lepro2022} for details of how the polarization rates were calculated.}
\label{fig:1}
\end{figure*}

\section*{Results}
\paragraph*{Large-scale transport is suppressed with increasing cargo size}
In Fig.~\ref{fig:1}(a), a microscopy image of a single amoeboid cell carrying a spherical microcargo is displayed, along with the tracks of the cell and the cargo particle shown in red and blue, respectively (see Movie~1 in the Supporting Material).
A corresponding cartoon of this biohybrid microtransporter and the spatial coordinates of the system are illustrated in Fig.~\ref{fig:1}(b).
We have previously shown in Ref.~\cite{Lepro2022} that the dynamics of this cell-cargo system exhibits two regimes, (1)~an idling rest state, where the cargo particle dwells at a constant equilibrium distance~$l_0$ from the cell center and performs circling movements around the cell, and (2)~intermittent transition phases during which the cell passes underneath the cargo and continues moving persistently for a while~[cf.~Fig.~\ref{fig:1}(c) for a cell trajectory shown in the frame of reference of the cargo, where episodes of circling motion and intermittent transitions can be clearly distinguished].
The transitions are initiated by bursts in cell polarity towards the cargo particle, most likely triggered by the mechanical impact of the cargo; their duration seems related to the intrinsic lifetime of cell polarity~\cite{Lepro2022}.
This pattern is also reflected in the temporal dynamics of the distance~$l(t)$ between the cell and cargo centers of mass, displayed in Fig.~\ref{fig:1}(d) for a cargo particle with a radius of~$23\,\mbox{\textmu m}$, where spikes of values in~$l$ reaching below the equilibrium distance (dashed red line) represent transition events.
The two dynamical regimes are also reflected in the histogram of~$l$-values taken over the experimentally recorded time series.
The main contribution to the histogram is due to fluctuations around~$l_0$ during the rest state, while the transitions contribute a second smaller peak at lower values of~$l$, resulting in an asymmetric histogram shape~[see Fig.~\ref{fig:1}(e)].

A similar pattern was observed for all recorded particle radii ranging from~$5$ to~$62~\mbox{\textmu m}$.
Whereas, the equilibrium distance~$l_0$ of the rest state increases with increasing particle size, the polarization rate~$\lambda$, at which transitions occur, decreases~\cite{Lepro2022}.
Previously, we have proposed an active particle model to account for the specific features of this intermittent colloid dynamics driven by a cell~\cite{Lepro2022}.
In particular, this modeling approach enabled us to calculate the long-time diffusion coefficient~$\mathcal{D}$ of the colloid as a function of the polarization rate~$\lambda$; in the absence of polarity bursts~($\lambda=0$), active transport vanishes and the diffusivity of the microtransporter decays to the value of the idling rest state.
For the present study, we extended our previous dataset to particles with a radius of~$62~\mbox{\textmu m}$ and extrapolated the decreasing polarization rate as a function of increasing particle radius to estimate the limiting particle size for which the polarization rate decays to zero, see Fig.~\ref{fig:1}(f).
From this estimate, we conclude that no transitions occur~--~thus, phases of persistent, polar movement will be absent~--~for particles with a radius that is larger than~$85~\mbox{\textmu m}$ (examples of cellular trucks with the corresponding cargo sizes can be seen in Movies~1-6).
In the absence of transitions, the active large-scale transport vanishes, reducing the dynamics of the system to circling of the cargo around the cell.
Therefore, the long-time diffusivity of the cellular truck should be determined by the diffusion coefficient of the carrier cell alone.

\begin{figure*}[t]
\centering
\includegraphics[width=.8\textwidth]{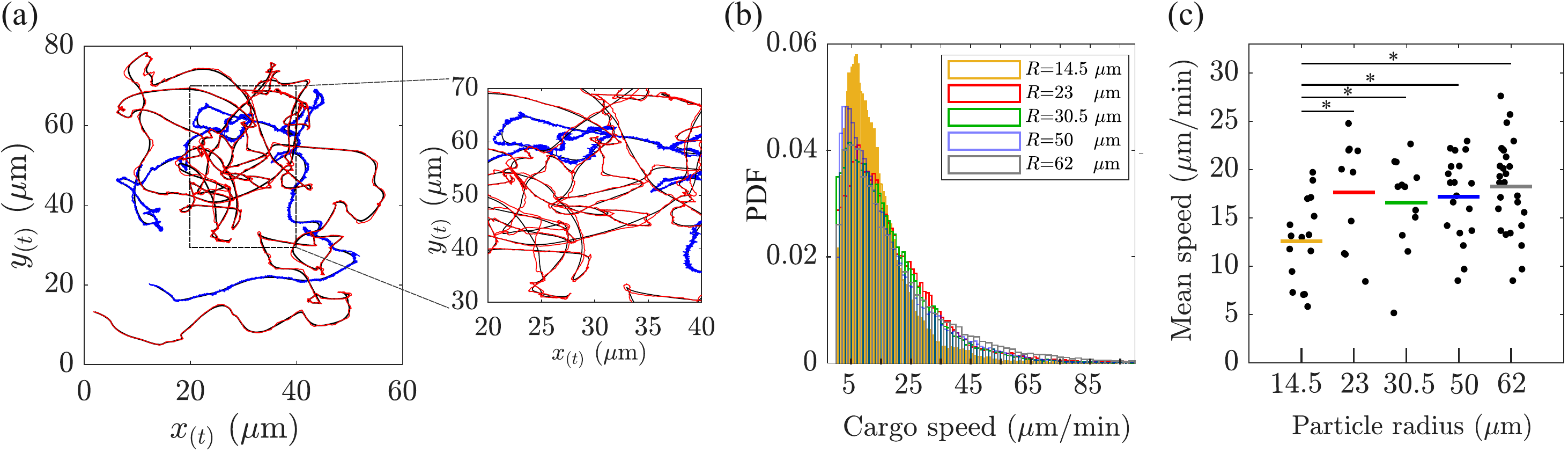}
\vspace{-0.2cm}
\caption{\fighead{Statistics of instantaneous cargo speeds.} (a)~Comparison between the original tracked positions of the cell (red) and the cargo (blue), with the smoothed trajectories (overlaid black lines). The smoothing windows~(moving average) are 150~frames (30~s) for the cell and 43~frames (8~s) for the cargo, respectively. (b)~Cargo speed distribution for each group of measured particle sizes. Only the particle group with~$R=14.5$~\textmu m (yellow distribution) shows a pronounced peak at lower speed values; others are practically indistinguishable. (c)~Mean cargo speed for each trajectory depicted as black data points that belong to the particle radii of~$R=14.5$~\textmu m ($n=16$),~$R=23$~\textmu m ($n=10$),~$R=30.5$~\textmu m ($n=12$),~$R=50$~\textmu m ($n=19$) and~$R=62$~\textmu m ($n=27$). The color-coded lines represent the averaged mean speeds for each group of particles (color-code is identical to the panel~b). A two-sample Kolmogorov-Smirnov test reveals that the speed statistics of cargo particles with a radius of~$R=14.5\,\mbox{\textmu m}$ is significantly different from larger particles~(significance level:~$\alpha = 0.05$), indicated by a star~(\text{*}); others are practically identical, i.e.~the speed statistics of cargoes does not significantly depend on the cargo radius for $R \ge 23 \, \mbox{\textmu m}$. }    
\label{fig:2}
\end{figure*}

\paragraph*{Cargo speed does not decrease with increasing cargo radius for intermediate cargo sizes}
In what follows, we will concentrate on the forces that the cell exerts on the cargo while moving it.
The force estimates do not depend on a non-zero polarization rate, as the cell exerts active forces onto the cargo also during the rest state, resulting in the characteristic circling motion around the cell.
We estimated the active force from the cargo speed by taking it to be approximately equal to the drag force that the cargo would experience when being displaced by the cell in a surrounding open, viscous medium:~the drag force on a sphere moving in a viscous fluid at low Reynolds number is given by Stokes' law,~$F = 6 \pi \eta R v$, where~$R$ is the radius of the sphere,~$\eta$ the viscosity of medium (here, taken to be equal to the viscosity of water at~$20^{\circ}$), and~$v$ is the speed of the sphere.

The force estimate thus depends on the instantaneous speed of the cargo.
To resolve also peaks in the active force that occurred during the microtransport, we performed recordings of the microtransport process with a temporal resolution of up to~$\Delta t= 0.2 \, \mbox{s}$, which is much shorter than the typical timescale of cargo motion.
The positions of the cell and the cargo were defined as the centers of mass of the connected identified regions, determined in every time frame through image segmentation (see Methods for details).
The instantaneous speeds of the cells and cargo particles were then calculated by finite differences.
Note that imaging noise and finite pixel resolution led to small fluctuations in boundary detection and center of mass calculation between consecutive frames.
This resulted in small errors in the displacements, which were amplified to large speed fluctuations by the small time step.
In order to avoid this artefact, trajectories were smoothed by moving averages prior to calculating the speed values.
The smoothing window was determined for every track individually from the correlation time of the original velocities as explained in the Methods Section.
Fig.~\ref{fig:2}(a) shows an example of the original cell (blue) and cargo tracks (red), in comparison to the smoothed trajectories (overlaid in black).

In Fig.~\ref{fig:2}(b), the speed distributions for cargo particles with radii ranging from~$14.5~\mbox{\textmu m}$ to~$62~\mbox{\textmu m}$ are shown.
They exhibit an asymmetric shape with a pronounced peak at speed values around 10~\textmu m/min and a tail ranging up to speeds of about 100~\textmu m/min.
Except for the speed distribution of the 14.5~\textmu m particles that displays a more pronounced peak and a more rapid decay, all other distributions closely overlap.
This is also reflected in the mean speed values that are similar for all cargo sizes, except for the 14.5~\textmu m particles that move at significantly smaller average speeds, see Fig.~\ref{fig:2}(c).
For particles with a radius of~$23~\mbox{\textmu m}$ and larger, a mean speed of transport equal to~$18\pm4$~\textmu m/min was observed, notably fairly independent of the particle radius~$R$.

\begin{figure}[t]
\centering
\includegraphics[width=\columnwidth]{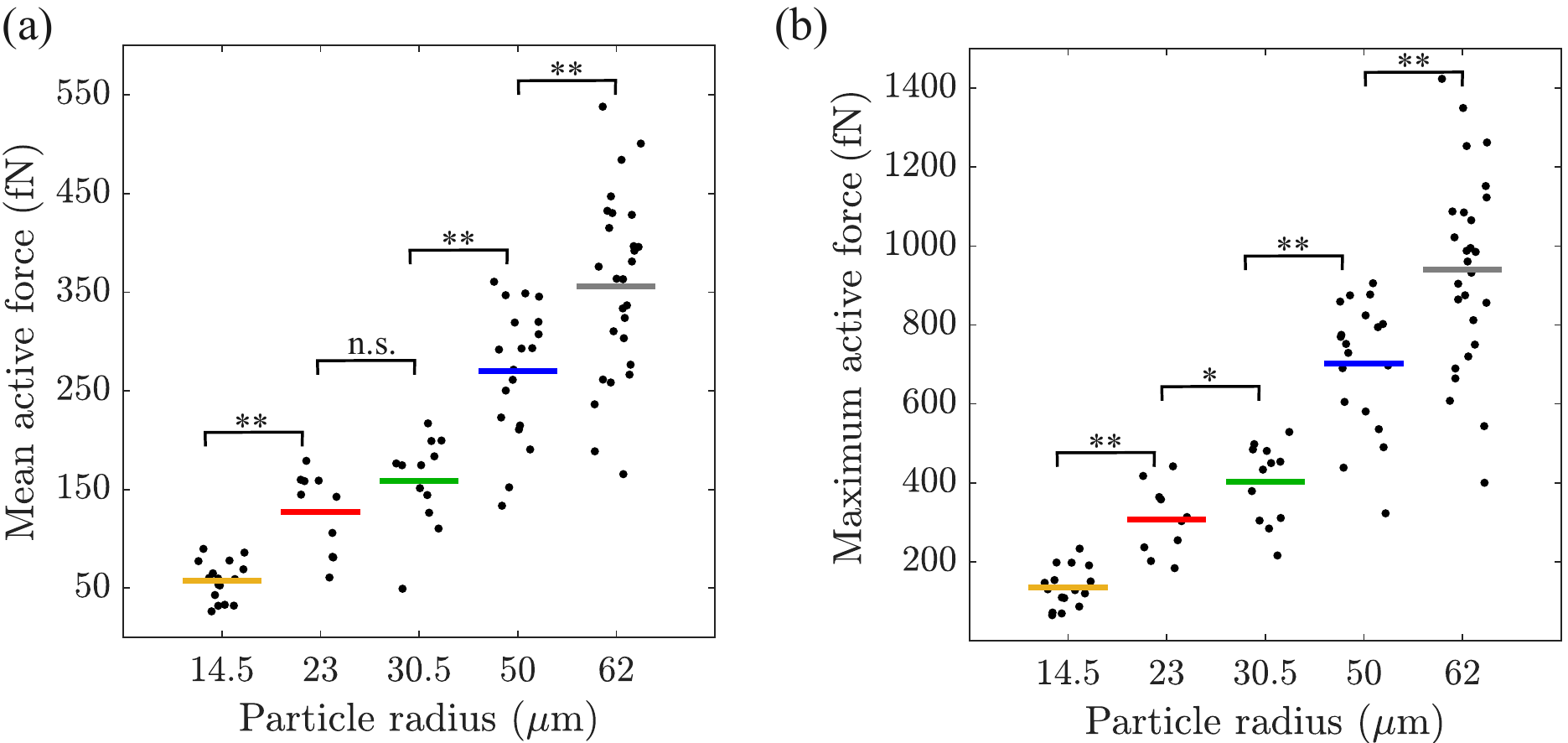}
\vspace{-0.2cm}
\caption{\fighead{Active forces exerted on the cargo.} (a)~On average, the active force~$f$ increases as a function of particle radius. Black data points represent the mean active force measured for each trajectory. The color-coded lines indicate the averaged mean force~$f$ for each group of particles. (b)~Black data points represent the maximum force (95\ts{th} percentile) recorded for every single trajectory. The color-coded lines represent the averaged maximum forces for each group of the particles. Brackets with one~(\text{*}) or two stars~(\text{**}) indicate a statistically significant increase of the force~(Kolmogorov-Smirnov with significance levels~$\alpha = 0.05$ and~$\alpha = 10^{-3}$, respectively; \textit{n.s.} stands for \textit{not significant}). }
\label{fig:3}
\end{figure}

\paragraph*{Only femto-Newton forces are required to move the cargo particles}
From the cargo speeds, we calculated the active forces that are necessary to displace the cargo particles based on Stokes' law.
As we found similar speeds for cargo radii between 23 and 62~\textmu m, the force according to Stokes' law increased with particle size.
We found mean forces ranging from 57~fN for 14.5~\textmu m particles to 356~fN for 62~\textmu m particles, see Fig.~\ref{fig:3}(a).
To estimate the maximum force applied to the cargo, we considered those episodes of transport with the highest (95th percentile) speeds.
On average, the maximum force also increased with particle size and reached values of up to 0.94~pN applied to particles with a radius of~$62~\mbox{\textmu m}$, see Fig.~\ref{fig:3}(b).

How will the cargo speeds and the active force applied by the carrier cell evolve for larger cargo particles? 
Unfortunately, the acquisition of reliable data in statistically sufficient amounts became more and more difficult with increasing cargo size; for particles with a radius of more than~$62~\mbox{\textmu m}$, it turned out to be practically impossible:~due to the large dimensions of the cargo in these cases, longer time series, where a single cell interacts with one cargo particle only, are difficult to capture.
Furthermore, it has often remained unclear in these situations, whether a neighboring cell is in physical contact with the cargo particle or not when very small colloid displacements were recorded.
In order to find out how the active force behaves for larger cargo sizes, we thus have to rely on modeling assumptions to extrapolate from the regime of our experimental observations to the speeds and corresponding forces that are expected for larger particles.

\begin{figure*}[t]
\centering
\includegraphics[width=0.8\textwidth]{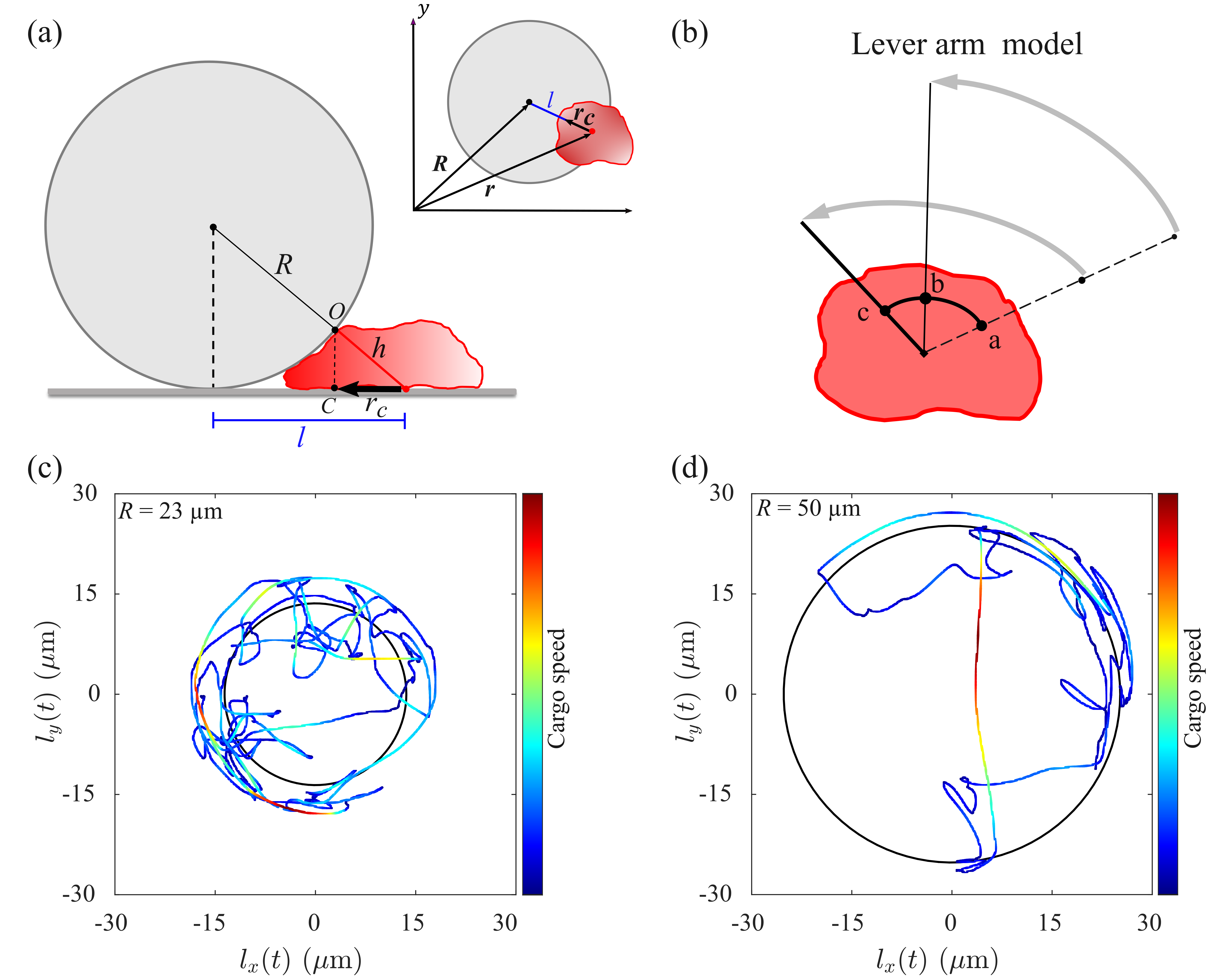}
\vspace{-0.2cm}
\caption{\fighead{Geometry of the system and the lever arm model.} (a)~Cross-section~(side view) of the cellular truck, exclusive to episodes of the idling state when the particle is in contact with the substrate. The cell is shown in red and the cargo in gray. The distance between the cell and the cargo centers of mass on the cell-substrate contact plane is denoted by~$l$; the cargo radius is~$R$. We denote further~$O$ as the point where the line that connects the center of mass of the cell-substrate contact area with the center of the spherical cargo particle intersects the cargo surface. The projection of this point on the substrate defines the contact point,~$C$, with~$\boldsymbol{r}_{c}$ describing its position vector with respect to the cell. The top view of the system is shown in the figure inset, cf.~Fig.~\ref{fig:1}(b). (b)~The lever arm model: The diagram explains the displacement of the cargo at the level of the cell-cargo contact point. The cartoon compares the displacement of two particles of different sizes in the laboratory frame within a fixed time interval. The larger particle tends to be further away from the cell~[larger equilibrium distance~$l_0$, cf.~Eq.~\eqref{eq:fit}]. The two gray arcs have the same length~--~however, the cell-cargo contact point at the level of the cell has moved a smaller distance~(from a to b) to displace the larger cargo in comparison to the smaller cargo, where the contact point has moved from a to c within the same time interval in order to achieve equal cargo displacements. In short, the speed of the cell-cargo contact point decreases as a function of the radius~$R$ but the recorded colloid speed in the lab frame remains unchanged. In panels (c) and (d), the cargo trajectories are shown in the frame of reference of the cell for a particle with~$R=23\, \mbox{\textmu m}$ and~$R=50\, \mbox{\textmu m}$, respectively; the color-code represent the magnitude of the cargo speed from blue (lowest) to red (highest). The solid black circle marks the equilibrium distance. }
\label{fig:4}
\end{figure*}

\paragraph*{Geometrical model for the speed of the cell-cargo contact point}
Our experimental results showed that cells displace cargoes of very different sizes, ranging from a radius of~$14.5~\mbox{\textmu m}$ to~$62~\mbox{\textmu m}$, with similar speeds.
According to Stokes' law, a cell thus invests more power on larger cargoes to achieve similar displacement speeds as for smaller cargoes.
Here, we propose a simplified geometrical model that we refer to as the ``lever arm model'' to interpret our findings in terms of the cell-cargo contact point, which will allow us to predict the cargo speed and active force beyond the experimentally accessible regime of cargo sizes.
In particular, we will estimate the maximum active force a cell exerts on a spherical cargo particle and the stalling point of active transport.

The concept of the lever arm model is based on the simplifying idea of reducing the complex and extended cell-cargo contact area to a single contact point.
Fig.~\ref{fig:4}(a) shows the overall geometry of the microtransport system in a side view; the corresponding top view can be seen in the inset.
We define the cell-cargo contact point as the location, where the line that connects the center of mass of the cell-substrate contact area with the center of the spherical cargo particle, intersects the cargo surface, marked as~$O$ in Fig.~\ref{fig:4}(a).
Its projection~$C$ into the cell-substrate plane lies on the line that connects the centers of mass of cell and cargo in our two-dimensional microscopy images~[vector~$\boldsymbol{l}(t)$], cf.~Fig.~\ref{fig:4}(a).
Assuming that the cargo particle is in contact with the substrate surface, we can calculate the position~$\boldsymbol{r}_{c}(t)$ of the projected cell-cargo contact point~$C$ as seen from the center of mass of the cell, using the positions~$\boldsymbol{r}(t)$ and~$\boldsymbol{R}(t)$ of the cell and cargo, respectively, in our two-dimensional microscopy images.
The position of the projected cell-cargo contact point is then given by
\begin{equation}
    \boldsymbol{r}_{c}(t) = \frac{l_{c}(t)}{l(t)} \cdot \Big [ \boldsymbol{R}(t) - \boldsymbol{r}(t) \Big ] 
    , \label{eq:cp_position}
\end{equation}
with
\begin{equation}
    l_{c}(t)=l(t)-\frac{R\,l(t)}{\sqrt{R^2+l^2(t)}}
    \label{eq:rc_abs}
\end{equation}
denoting the distance from the center of mass of the cell-substrate contact area to the projected cell-cargo contact point~$C$, depending on the cell-cargo distance~$l(t)$ and the radius~$R$ of the cargo particle.
As movements of the cell-cargo contact point~$O$ in vertical direction will be much smaller than the lateral movements, reflected by the circling motion of the cargo around the cell, we will estimate the speed of the cell-cargo contact point from the speed of its projection~$C$ in the cell-substrate plane.

Note that Eq.~\eqref{eq:cp_position} critically depends on the condition that the cargo is in contact with the substrate.
For larger cargo sizes, transitions become rare, see Fig.~\ref{fig:1}(f).
In this regime, the cargo dynamics is limited to circular motion around cell, so that we can safely assume that the cargo remains close to the substrate.
For smaller cargoes, however, this is not necessarily the case.
To estimate the contact point speed, we therefore rely only on the cargo speed values taken from those episodes of the data for which the condition~$l(t) \geq l_{0}$ is fulfilled, i.e.~for which the cell-cargo distance~$l(t)$ is equal or larger than the equilibrium distance~$l_0$~(during the rest state), so that we can assume that the cargo is in contact with the substrate, thereby excluding transitions from the data analysis.
The data analysis revealed that force maxima appear not only during the transitions but also during the resting state, when the cargo particle circles around the cell at a fixed distance~[see Fig.~\ref{fig:4}(c,d), where the trajectory of the cargo, seen from the frame of reference of the cell, is shown with a color-code corresponding to the cargo speed].
We are thus confident that maxima of the active force can be also reliably estimated from cargo trajectories excluding the transition events.

\paragraph*{Speed of the cell-cargo contact point decreases with cargo size predicting an upper size limit for active transport}
Based on the geometrical model introduced above, we now ask how the speed of the cell-cargo contact point will evolve for larger cargoes and what limiting cargo size the cell will fail to move.
Finding these limits will also provide an estimate of the maximum force that a single agent cell applies to the cargo during transport in an isotropic viscous fluid environment.

\begin{figure*}[t]
\centering
\includegraphics[width=.8\textwidth]{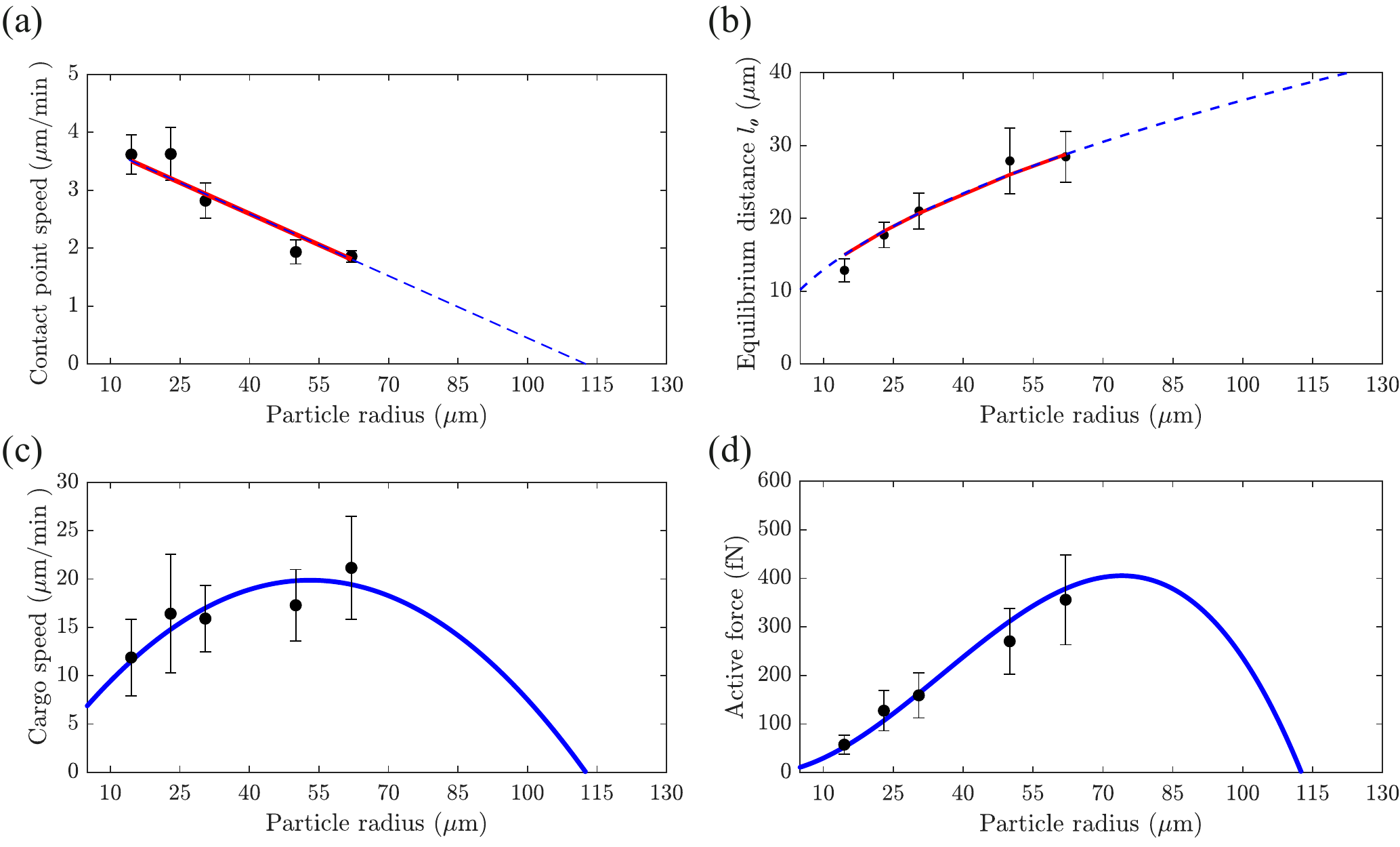}
\vspace{-0.2cm}
\caption{\fighead{Results of the geometrical model.} (a)~The descending contact point speed as a function of cargo radius. For each particle size, the black data points represent the median of the contact point speeds. The error bars are the standard error of the mean~(2$\sigma$ interval); see Fig.~\ref{fig:S3} in Supporting Material for details of the contact point speed statistics. The red solid line shows a linear fit to the data points. The particle radius at which the contact point reaches zero values is $113 \, \mbox{\textmu m}$~($1\sigma$-confidence interval:~$[104\,\mbox{\textmu m},124\,\mbox{\textmu m}]$, estimated from the uncertainties of the inferred fit parameters). (b)~The mean equilibrium distance~$l_{0}$ (black points) and the law fit to the measured data (red solid line) using Eq.~\eqref{eq:fit}. (c)~The speed of the cargo estimated for the full range of particle sizes. For each particle size, the black points are the mean cargo speeds derived from isolated idling states of all trajectories. The solid blue line is the mean cargo speed predicted by the geometrical model. (d)~Using the cargo speeds from the geometrical model, the mean active force is estimated for the full range of particle sizes. The black points are data of mean active forces (identical to Fig.~\ref{fig:3}). The solid blue line with a peak force at approximately $400\, \mbox{fN}$ is the mean active force estimated from the geometrical model. The error bars of the data points in panels~(b) and (d) represent standard deviations. } 
\label{fig:5}
\end{figure*}

After excluding the transition periods as described above, we determined the position~$\boldsymbol{r}_c(t)$ of the cell-cargo contact point as seen from the center of mass of the cell according to Eq.~\eqref{eq:cp_position}.
In the laboratory frame of reference, the position of the cell-cargo contact point is thus given by~$\boldsymbol{R}_c(t)=\boldsymbol{r}(t)+\boldsymbol{r}_c(t)$ and its speed by the time derivative of~$\boldsymbol{R}_c(t)$.
Since the speed of the cell is much smaller than the speed of the cargo particle (by about a factor of 5 on average, see Fig.~\ref{fig:S2}), we approximate the speed of the contact point by~$\dot{\boldsymbol{R}}_c(t) \approx \dot{\boldsymbol{r}}_c(t)$, thus assuming that the cell remains stationary at the timescale of interest and the cargo circles around it.
The resulting contact point speeds for different cargo sizes can be seen in Fig.~\ref{fig:5}(a).
While the speed of the cargo remained roughly constant for intermediate particle sizes, the contact point speed decreases~(see the Supporting Material for details of the contact point speed statistics, in particular Fig.~\ref{fig:S3}).
This can be understood as a consequence of the geometry of the system that is represented in a simplified fashion by our lever arm model, see Fig.~\ref{fig:4}(b).
Spheres that are displaced at similar speeds along circular trajectories around the cell will exhibit a decreasing cell-cargo contact point speed for increasing radius~$R$ of the spherical cargo particles.

We applied linear regression to the contact point speed as a function of the particle radius and extrapolated the fit function to find a critical cargo radius of 113~\textmu m, where the contact point speed decreased to zero.
From the uncertainty of the fitting parameters, we expect the critical cargo radius to fall into the range between 104 and 124~\textmu m.
Note that this critical radius, where movement of the cargo is expected to stall, is an averaged quantity.
Due to cell to cell variability, this limit may vary considerably between individual cells.

\paragraph*{Estimate of cargo speed and active force beyond the experimentally accessible regime}

Having determined the full range of cargo sizes that can be moved by an amoeboid carrier cell, we can now estimate the cargo speeds and the active forces from the contact point speed also for larger cargoes outside the experimentally accessible range.
For this purpose, we use the established relation of the measured cargo dynamics~$\boldsymbol{R}(t)$ and the contact point~$\boldsymbol{r}_c(t)$ via Eqs.~\eqref{eq:cp_position}-\eqref{eq:rc_abs}; as argued before, the cell speed can approximately be neglected~--~the cell is considered non-motile at the timescales of interest.
Note that for cargo sizes outside the regime that we can analyze in our experiments, the time series~$l(t)$ of the cell-cargo distance is no longer available.
That is why we approximate~$l(t)$ in Eqs.~\eqref{eq:cp_position}-\eqref{eq:rc_abs} by the equilibrium distance~$l_0=l_0(R)$, which depends on the cargo radius.
For the experimentally accessible cargo sizes, the equilibrium distance~$l_0$ increases with increasing cargo radius~$R$, see Fig.~\ref{fig:5}(b).
We fit the dependence of~$l_0$ on~$R$ using the model function
\begin{equation}
    l_0(R)=\sqrt{\left (h+R \right )^2 - R^2} \, = \sqrt{h^2+2Rh} \, , \label{eq:fit}
\end{equation}
which was inspired by the geometry of the system, cf.~the right-angled triangle in Fig.~\ref{fig:4}(a) with hypotenuse length~$h+R$ and the legs~$\{ R , l_0\}$.
Note that this fit to the experimental data yields an estimate of~$h=6.4$~\textmu m, which is a reasonable number given the typical dimensions of a {\it D.~discoideum} cell~\cite{Tanaka2020,Hrning2019}.

Extrapolating the contact point speed~[Fig.~\ref{fig:5}(a)] and the fit function~[Eq.~\eqref{eq:fit}] to larger cargo radii~[Fig.~\ref{fig:5}(b)], we obtain estimates of the average cargo speeds by differentiating Eq.~\eqref{eq:cp_position} with respect to time.
In Fig.~\ref{fig:5}(c), the resulting cargo speeds are displayed as a function of the radius~$R$ (blue line).
For smaller cargo sizes, the speed increases and reaches a maximum of 20~\textmu m/min for cargo radii of around 55~\textmu m.
Towards larger sizes, the speed then decreases and drops to zero at the critical radius of 113~\textmu m.
This is in good agreement with the experimentally measured speeds during the rest phase, displayed as black data points in Fig.~\ref{fig:5}(c).
The data, however, shows strong fluctuations and is limited to sizes up to a radius of 62~\textmu m.
The speeds of larger cargoes are only accessible based on the proposed geometrical lever arm model.

Finally, we also estimated the corresponding active forces according to Stokes' law based on measurements of the cargo speeds.
As shown in Fig.~\ref{fig:5}(d), the force increases for smaller radii and goes through a maximum, before dropping to zero at the limiting cargo radius of 113~\textmu m.
The peak value of 0.4~pN is reached for a particle radius of 74~\textmu m.
As already indicated above, this maximum is an average value.
Depending on the individual cell, peak force of around 1~pN can be observed over short periods of time~[Fig.~\ref{fig:3}(b)].

\begin{figure*}[t]
\centering
\includegraphics[width=.8\textwidth]{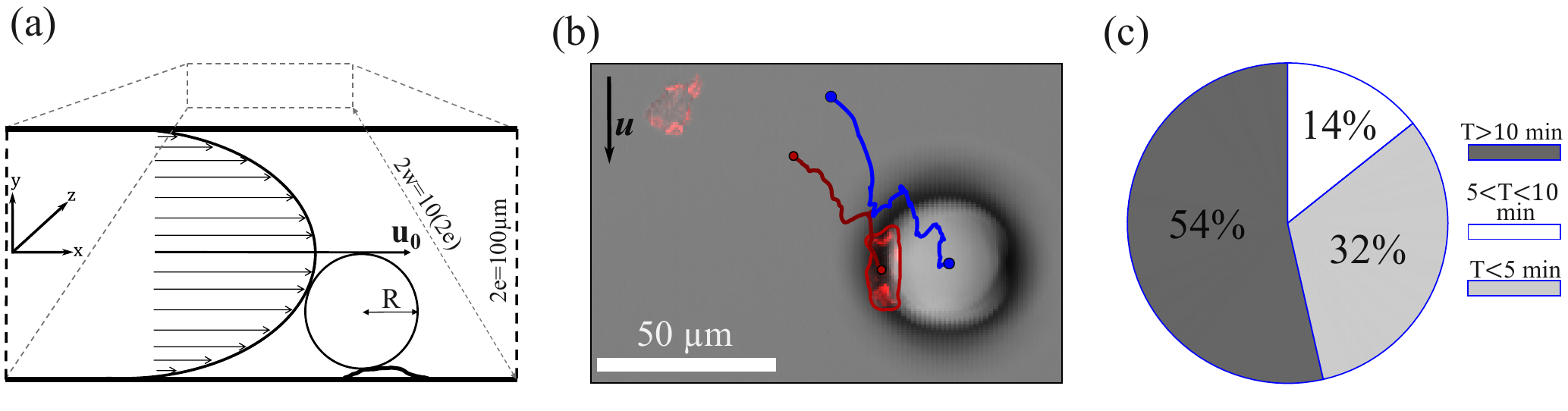}
\vspace{-0.2cm}
\caption{\fighead{Cellular truck under a constant drag force.} (a)~Cross-section of the micro-channel including main geometrical quantities and the parabolic flow profile;~$u_{0}$ is the maximum flow speed at the center of the channel~(geometry: $e=50\, \mbox{\textmu m}$, $w=0.5 \, \mbox{mm}$). The cell is in contact with a particle of radius~$23~\mbox{\textmu m}$, covering up to the half-height of the channel. 
(b)~A microscope image of the cell (red) and cargo (gray sphere) under flow conditions. The tracks of the cell (in red) and the cargo (in blue) show the displacement of the entire system downstream (Movie~7). The black arrow shows the direction of the flow. 
(c)~The pie chart characterizes the total percentage of the loaded cells under the critical flow that withhold to the cargo over $10 \, \mbox{min}$ (54\%,~$n=15$), between $2 \, \mbox{min}$ to $5 \, \mbox{min}$ (14\%,~$n=4$), or lose the cargo shortly after the flow insertion (32\%,~$n=9$). The flow rate is $3 \, \mbox{\textmu l/min}$.}
\label{fig:6}
\end{figure*}

\paragraph*{Cell-cargo interaction under a constant external pulling force}
So far, we have focused on amoeboid microtransport on an open flat substrate in a uniform viscous fluid medium at rest.
However, a carrier cell will typically experience more complex environments, where friction forces due to geometrical confinement or fluid flow may additionally affect the cargo particle.
As biological cells including {\it D.~discoideum} are mechanoresponsive~\cite{Roth2015,Nagel2014,Dalous2008}, we expect that the active forces exerted by the carrier cell may change if additional external forces are acting on the cargo particle.
To provide a first estimate of how external conditions may affect the force generation of the carrier cell, we exposed amoebae that were loaded with a spherical cargo particle to a constant drag force generated by fluid flow in a microfluidic device, see Fig.~\ref{fig:6}(a,b) and Movie~7 in the Supporting Material.

A schematic of the rectangular flow chamber with half height~$e$ and half width~$w$~(much larger than the height:~$w \gg e$) is depicted in Fig.~\ref{fig:6}(a).
Since the Reynolds number is low, inertial effects can be neglected.
The flow between parallel plates then exhibits a parabolic profile,
\begin{equation}
    u(y)=u_{0} \! \left ( 1-\frac{y^2}{e^2} \right ) \quad\mbox{with}\quad
    u_{0}=\frac{dp}{dx}\cdot\frac{e^2}{2\eta} \, , \label{eq:flow}
\end{equation}
where~$u_{0}$ is the flow speed at the center of the channel,~$\eta$ is the viscosity of the medium (here, taken to be equal to the viscosity of water at~$20^{\circ}$C) and~$dp/dx$ is the pressure gradient along the length of the channel~\cite{1928,Holmes1968}.
In our experiments, we used particles with a radius of 23~\textmu m.
We gradually increased the flow rate and thereby the drag force on the cargo particle to the point where about half of the cells lost their cargo, indicating that we approached the critical force required to rupture the adhesive bond between the cell and the cargo particle.
At a flow rate of 3~\textmu l/min~--~corresponding to a peak velocity of 0.75~mm/s and a pressure gradient of 0.6~Pa/mm~--~we observed that~$54\%$ ($n=15$) of the cells maintained their adhesion to the cargo particle, while the remaining~$46\%$ ($n=13$) lost connection to the cargo either immediately or within 5 minutes after starting the fluid flow~[cf.~Fig.~\ref{fig:6}(c)].

\begin{figure*}[t]
\centering
\includegraphics[width=.8\textwidth]{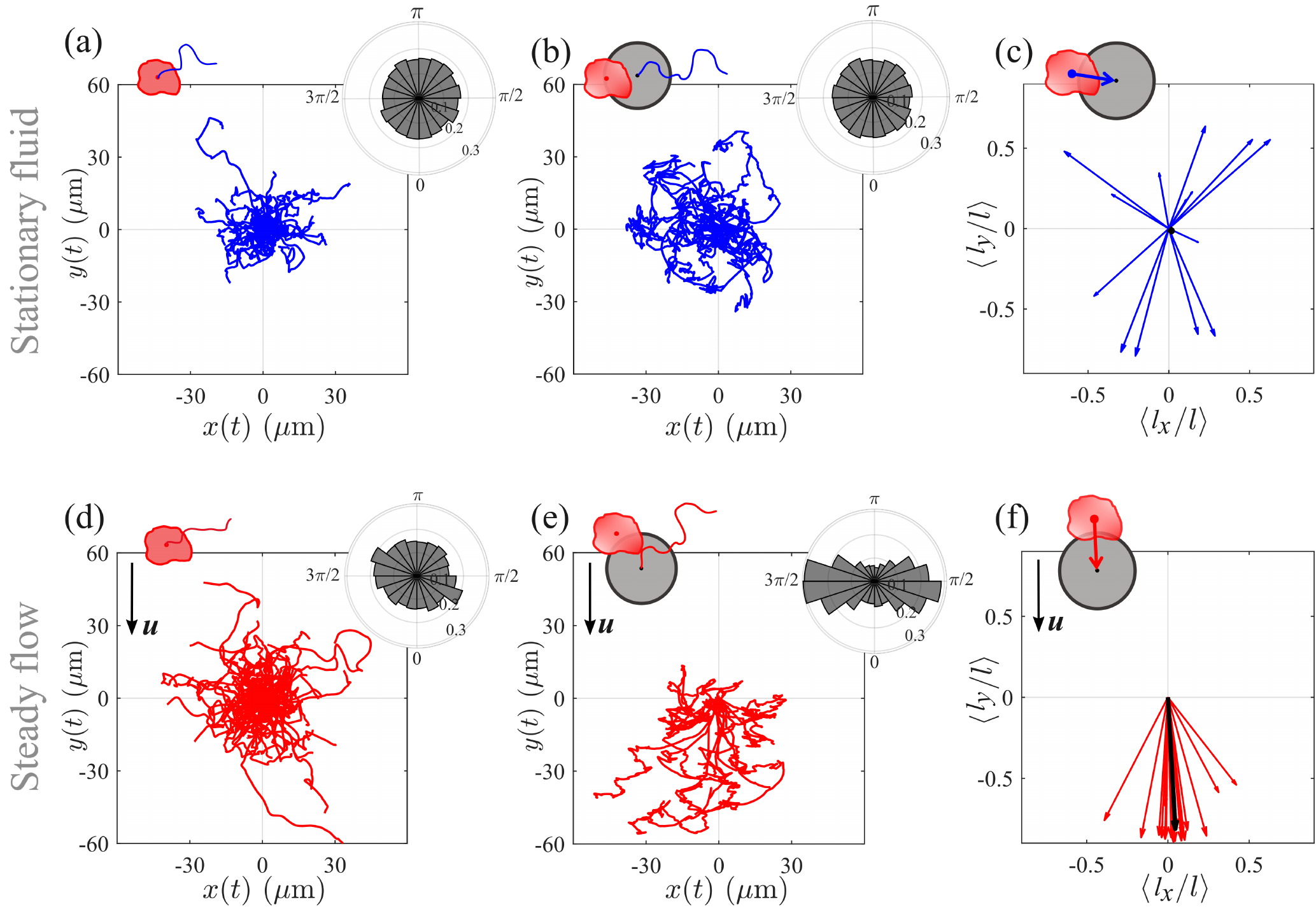}
\vspace{-0.2cm}
\caption{\fighead{Dynamics of unloaded cells and cellular trucks under flow conditions.} In panel (a), the trajectories of single cells~($n=77$) and in (b), the trajectories of cargoes from cellular trucks~($n=15$) are shown from a common origin in the absence of flow. In panel (d), the trajectories of single cells~($n=104$) and in (e), the trajectories of cargoes from cellular trucks~($n=15$) are shown from a common origin under a Poiseuille flow (flow rate of $3 \, \mbox{\textmu l/min}$). Individual cells show no sign of directional motion in both, stationary and flow conditions. In contrast, cellular trucks tend to move downstream on average under flow conditions as shown in panel~(e). The polar plots in (a,b) and (d,e) show the angle between the respective velocity vector and the flow direction. The polar plot~(e) suggests that the cell displaces the cargo mainly perpendicular to the flow direction. The arrows in panels~(c) and~(f) show the mean of the normalized displacement vector~$\boldsymbol{l} / l$, where~$\boldsymbol{l} = \boldsymbol{R}- \boldsymbol{r}$, i.e.~the cargo position with respect to the cell for a stationary fluid~[panel~(c)] and under flow~[panel~(f)]; the overlaid black arrow is the average. Whereas a cell can place the cargo at arbitrary positions in the absence of flow, the particle tends to be placed downstream under flow conditions. }
\label{fig:7}
\end{figure*}

To ensure that the observed effects are only due to the drag force acting on the cargo particle, the wall shear stress should remain below the critical value of 0.8~Pa above which {\it D.~discoideum} cells start exhibiting shear induced directional responses~\cite{Dcav2002,Decave2003,Dalous2008}.
For the critical flow rate of 3~\textmu l/min, the wall shear stress is 30~mPa, remaining indeed below the threshold value of 0.8~Pa.
Consequently, comparing the tracks of single cells without cargo in stationary liquid (blue tracks) and under flow (red tracks), no directional movement was detected, see Fig.~\ref{fig:7}(a) and~(d).
Only a moderately enhanced but isotropic spreading was observed in the presence of fluid shear stress.

In contrast, a clear directional bias was observed in the presence of a fluid flow for cells that carry a cargo particle, see Fig.~\ref{fig:6}(b) for an example:~the entire cellular truck was pulled downstream as a consequence of the drag force acting on the cargo particle~[see Fig.~\ref{fig:7}(e) for several examples of cargo trajectories under fluid flow].
Trucks in the absence of fluid flow, on the other hand, spread isotropically, as shown in Fig.~\ref{fig:7}(b).
Moreover, we found that the cargo particle is oriented exclusively in downstream direction with respect to the cell under fluid flow, see the red arrows in Fig.~\ref{fig:7}(f), while the cell can position the cargo in any direction at its periphery in the isotropic environment of a resting fluid, see the blue arrows in Fig.~\ref{fig:7}(c).

In the microfluidic setup, the cargo radius is comparable to the channel height. 
Taking into account the top and bottom channel boundaries, the hydrodynamic drag force~$F$ acting on the cargo particle is not exactly following Stokes' law, which is valid only far from boundaries, but there is a correction factor~$f$ that depends on the radius~$R$ of the sphere with respect to the channel height as well the distance of the sphere from the walls of the channel~\cite{Jones2004,Cichocki1998,Cichocki2000}:~$F=6\pi \eta f R u_{0}$.
Following Refs.~\cite{Cichocki1998,Jones2004}, we rely on an approximation that represents the analytical solution of the Stokes equation in the two-wall geometry as a superposition of two individual walls.
Assuming that the cargo resides on top of the cell~(thereby setting a minimal distance of the cargo from the wall), the geometric-dependent correction factor for our experimental setup is approximately~$f \approx 1.54$.
As a result, we estimate the critical drag force to be~$F=6\pi \eta f R u_{0} \approx 0.5 \, \mbox{nN}$.

\section*{Discussion and Conclusion}

Here, we explored the active forces involved in cell-driven microtransport by monitoring the movement of spherical cargo particles and estimated the corresponding drag forces that act on the cargo particles based on their speed in an isotropic viscous fluid environment.
The speed of transport remained approximately constant for all tested particle sizes except for the smallest particles of radius~$R=14.5$~\textmu m, where lower speeds were observed.
This is in line with a lever arm model based on the geometry of the system.
Another reason for a decreased speed for smaller particles could be the phagocytic capacity of {\it D.~discoideum} cells~\cite{Dunn2018,Clarke2010,Xu2021}:~while the curvature of larger particles is not sufficient to trigger phagocytosis, particles with a radius of~$14.5~\mbox{\textmu m}$ may be small enough to stimulate attempts, albeit unsuccessful, to engulf the particles, resulting in reduced pseudopod formation, motility and cargo transport.
The similar speeds for larger particle sizes imply that cells generate active forces that increase with particle size, resulting in averaged forces up to 0.36~pN that were observed in our experiments.

We see two reasons for a dependence of the active forces on the cargo size.
Firstly, the cellular microenviroment, in particular the degree of confinement, strongly affects cell motility via mechanosensitive responses of the cytoskeletal activity and the cytosolic pressure~\cite{Yoshida2006,Pieuchot2018,Nagel2014,Srivastava2020,IshikawaAnkerhold2022,Krishnamoorthy2020,Roth2015,Sadjadi2022}.
In our cargo-transport situation, the cell is confined between the substrate and the surface of the cargo particle, so that particles of different radii expose the cells to geometrically different confinements and, thus, different mechanical stimuli.
This may trigger different confinement-induced responses, resulting in elevated active forces for larger cargo particles.
Note, however, that the speed of the cargo particle is not directly related to the speed of the carrier cell but additionally affected by shape changes and cytoskeletal activity at the dorsal cell cortex.
Secondly, larger particles experience larger drag forces and, thus, resist the cell-driven transport more strongly than smaller particles.
Also this may trigger more intense cytoskeletal activity resulting in larger active forces~--~an aspect that is also supported by our observations of transport under flow conditions, see below.

Even though we did not observe a decreasing speed of transport for larger particles, transport will eventually stall for very large cargo sizes, as the active forces that a single cell can generate must be limited.
Unfortunately, this regime is not accessible in our experimental setting.
For larger particle sizes, it becomes increasingly difficult to ensure that only one carrier cell is in contact with the cargo.
Moreover, it is difficult to decide whether a cell-cargo contact is actually established or not for large particles that are hardly displaced by single cells.
Given that we need a sufficient number of cargo trajectories to reliably estimate the speed and drag force on the cargo particle, data on cargo sizes beyond a radius of~$R=62$~\textmu m was not accessible in our experiments.
To close this gap, we proposed a simple geometrical model to estimate the position and speed of the cell-cargo contact point.
Together with our experimental data, this model shows that the speed of the contact point decreases with increasing particle size, allowing us to extrapolate towards larger cargoes and to estimate the critical cargo size, where cell-driven transport finally stalls.
Based on this approach, we predict that single motile {\it D.~discoideum} cells can move spherical particles up to radii of approximately~$113~\mbox{\textmu m}$.

Note that the maximum forces we observed did not exceed averaged values of 0.36~pN.
Also the forces predicted by our model for larger particles sizes beyond the experimentally accessible regime do not reach values larger than 0.40~pN.
Compared to other forces that are observed on the cellular scale, such as adhesion forces, cell-substrate traction forces, or even forces exerted by single molecular motors, these are very small values~\cite{DelanoAyari2010,Zimmermann2012,Bastounis2014,lvarezGonzlez2015,Kamprad2018,Srivastava2020}.
We thus conclude that it is not the Stokesian friction that imposes a size limit on the cargo particles that can be transported by single amoeboid cells.
Instead, we conjecture that the decreasing speed of the cell-cargo contact point may have geometrical reasons.
For small cargo radii the cell will experience a wedge-shaped confinement that stimulates and guides its migration in an attempt to maximize contact to available surfaces~\cite{Arcizet2012}, including intermittent bursts of polarization in the direction of the cargo particle~\cite{Lepro2022}.
With increasing particle radius, the difference in slope between the confining bottom and top surfaces will become less and less pronounced, resulting in less frequent polarity bursts, see Fig.~\ref{fig:1}(f), and a decreasing overall transport activity that is reflected by a decay of the cell-cargo contact point speed, see Fig.~\ref{fig:5}(a).

The behavior of cargo loaded cells under fluid flow, resulting in constant drag forces as high as 0.5~nN, is supporting our interpretation.
In our experiment, half of the population of cells carrying particles with a radius of~$23~\mbox{\textmu m}$ could resist this drag force for over 10~minutes.
In particular, among the cargo trajectories of this population that were mostly drifting downstream, we also observed short episodes, where the cargo was moved against the flow-induced drag force of 0.5~nN (see Movie~8).
This demonstrates that the active forces exerted by the cell can, at least shortly, exceed the Stokesian friction forces that arise when moving the cargo in a liquid at rest by three orders of magnitude.
We assume that these peak forces are triggered in response to the mechanical stimulus the cell experiences when the fluid flow is pulling the cargo particle.
For larger flow speeds, the resulting drag force will exceed the average cell-cargo adhesion force and detach the cargo particle from the cell in most cases.
The estimated cell-cargo adhesion strength is 2-4 fold smaller than the known cell adhesion forces to a glass substrate~\cite{Kamprad2018,Benoit2002}.
This is in line with our experimental observation that cells remain attached to the substrate even after loosing the cargo.

To conclude, we showed that single amoeboid cells are capable of transporting cargo particles significantly larger than their own body size exerting minuscule forces in the sub-piconewton range.
These forces can increase by several orders of magnitude if an external counter force is actively pulling on the cargo particle.
Our findings highlight the potentials and limits of amoeboid cells for designing autonomous biohybrid transport systems and for studying future applications of these systems when operating under more complex, real world conditions.
As amoeboid motility is common to many mammalian cell types, our findings will be relevant for putting biohybrid transport in medical applications into practice.



%



\section*{Acknowledgments}

S.S.P.~and C.B.~gratefully acknowledge funding by Deutsche Forschungsgemeinschaft (DFG) via project \textit{Sachbeihilfe BE 3978/3-3}.   
V.L.~and C.B.~acknowledge financial support via the IMPRS \textit{Multiscale Bio-Systems}. 
We thank Vedrana Fili\'{c}, Maja Marinovi\'{c}, and Igor Weber~(Rudjer Boskovic Institute, Zagreb, Croatia) for providing the Lifeact-mRFP encoding plasmid, and Kirsten Sachse as well as Maike Stange for technical support.

\section*{Author contributions}

S.S.P.~conducted experimental research; 
V.L.~contributed experimental data; 
R.G.~contributed to the modeling; 
S.S.P.~and C.B.~wrote the manuscript; R.G.~and V.L.~commented on the draft. 
C.B.~designed and supervised the project.

\section*{Competing interests}

The authors declare no competing interests.


\section*{Data availability}

The data that support the plots within this paper and other findings of this study are available from the corresponding author upon request.

\vspace{1em}
\noindent
\textbf{Correspondence and requests for materials} should be addressed to C.B.  

\vspace{1em}


\balancepage



\small{

\section*{Methods}

\vspace{-0.35cm}

\metparagraph*{Cell culturing.}

LifeAct-mRFP AX2 axenic {\it D.~discoideum} mutant cells were cultivated in tissue culture flasks (TC Flask T75 Standard, Sarstedt AG \& Co. KG, Nümbrecht, Germany) in a nutrient medium (HL5 medium including glucose supplemented with vitamins and micro-elements, Formedium Ltd., Norfolk, England) at 20~\textdegree C.  The medium was supplemented with a penicillin~(final concentration:~$100 \, \mbox{I.U.}/\mbox{ml}$) and streptomycin~(final concentration: $100 \, \mbox{\textmu g}/\mbox{ml}$) antibiotics mix~(CELLPURE\textregistered  Pen/Strep-PreMix, Carl Roth GmbH+Co. KG, Karlsruhe, Germany) and G418~(G418 disulfate ultrapure, VWR International, LLC.) as selection agent~(final concentration of $10 \, \mbox{\textmu g}/\mbox{ml}$). Prior to the first harvest of the cells, the spores were grown adherently to the glass bottom dish for 4 up to 7 days followed by a renewal of the medium every second day. After reaching over 50\% confluent monolayer, the cell suspension was diluted (ratio 1:200 of cell vs.~medium) at each medium renewal to avoid over-confluency. In addition, the entire cell culture was renewed every four weeks to avoid the accumulation of any undesired mutation arising from genetic drift.

\metparagraph*{Sample preparation.}

Monodispersed spherical particles (microParticles GmbH, Berlin, Germany) with a diameter range between 10~\textmu m to 214~\textmu m were stored in deionized water at 4-7~\textdegree C. Prior to the experiment, cells were harvested from the cell culture flask. The cell suspension was then diluted to obtain a cell count of roughly $5 \cdot 10^{4}$~cells$/\mbox{ml}$ for experiments with particles of up to 75~\textmu m and $3 \cdot 10^{4}$~cells$/\mbox{ml}$ for larger particle sizes. 1.5~ml of the cell suspension was then transferred into a culture dish (FluoroDishTM tissue culture dish with a cover glass bottom -- 35~mm, World Precision Instruments, Inc., Sarasota, Florida, USA). Cells sedimented and adhered to the bottom of the dish within $15 \, \mbox{min}$. Afterwards, 8-15~\textmu l of the particle suspension was added to the sample to achieve an approximate particle-cell ratio of 1:5. The sample was then gently shaken to achieve a uniform particle distribution. Before the imaging, the sample was kept at rest for another period of $15 \, \mbox{min}$.

Microfluidics: We used glass bottom channels with rectangular cross sections, purchased from ibidi\textregistered~(\textmu-Slide VI 0.1: 0.1~mm height, 1~mm width, ibidi GmbH, Martinsried, Germany). First, the channel was filled with 1.7~\textmu l of cell suspension containing $2 \cdot 10^{6}$~cells$/\mbox{ml}$. The sample was kept at rest for $15 \, \mbox{min}$ to ensure cell-substrate adhesion. Next, a few droplets of the dense particle suspension was added to the channel inlet. A 1~ml gas-tight microsyringe (Harvard Apparatus, Holliston, USA) was filled with cell culture medium and mounted on a PHD Ultra micropump (Harvard Apparatus, Holliston, USA), then gently connected to the channel inlet via PTFE tubing (FEP Tubing, 1/16~inch outside diameter, 0.03~inch inside diameter, IDEX HEALTH \& SCIENCE, USA). The small pressure resulting from connecting the tubes to the channel inlet was sufficient to push the particles into the channel. Accumulated cells and particles in the channel inlet as well as any trapped air bubbles were then removed by applying a gentle flow. The sample with all connected tubes was kept at rest for further $30 \, \mbox{min}$ to reach a stationary state. Subsequently, the medium was injected into the channel at a constant flow rate of $3 \, \mbox{\textmu l}/\mbox{min}$.

\metparagraph*{Imaging.}

Imaging was performed using confocal laser scanning microscopy (LSM 780, Zeiss, Oberkochen, Germany). The fluorophore mRFP, colocalized with the F-actin of the cytoskeleton of cells, was excited with a 561~nm laser for cell detection~(pinhole aperture of 1~Airy unit, 40x/64x objectives). The transmitted light was then band-pass filtered (585-727~nm) and collected by a photo-multiplier. The full spectrum transmitted light was also collected using a second acquisition channel where the bright-field images were generated as a result of discontinuous refractive indices. These images were then used for particle detection. The focal plane was adjusted to the height where the ventral surface of the cell meets the substrate and the lower section of the particles appeared as a bright spot surrounded by a dark ring. Images were acquired with a sampling time~(time interval) of~$\Delta t = 0.197 \, \mbox{s}$. Individual cellular trucks were recorded as long as possible, with a maximum of~$2 \, \mbox{h}$; the measurement time is limited by interruptions, such as collisions with neighboring particles, cell division or the interference with another cell.

For experiments in microfluidic channels, the sample was imaged at a frame rate of~$2 \, \mbox{fpm}$ as long as the cargo remained attached to the cell~(up to $40\, \mbox{min}$). The recording was stopped or discarded if floating cells bound to the cell-particle configuration of interest or if other interruptions occurred~(cf.~discussion above).

\metparagraph*{Image analysis.}

The image processing was performed using custom algorithms written in Matlab (R2021b, MathWorks, Natick, MA, USA).

Cell and particle segmentation was based on the images from the fluorescent channel collecting the emission signals of the labeled F-actin and the bright-field channel that collects the transmitted light, respectively. The image sequence was initially subjected to noise reduction using median filtering, followed by contrast enhancement protocols including a sequence of nonlinear histogram remappings. Subsequently, a threshold determined by the Otsu method~\cite{Otsu1979} was applied to the preprocessed images for binarization. The binarized images were then segmented followed by tracking of the resultant objects~\cite{crocker_methods_1996} based on the center of mass of segmented regions. In the case of cells, segmented boundaries were beforehand processed with an active contouring algorithm~\cite{ChenyangXu1998,Driscoll2012,LeproDissertation}; the resulting boundaries were used to determine the cell's center of mass. Note that we defined the cell and the cargo positions as the two-dimensional center of mass coordinates, derived from the connected components of binarized images.

\metparagraph*{Data analysis.} The complete statistical analysis was performed in Matlab (R2021b, MathWorks, Natick, MA, USA).

To avoid spurious fluctuations of the velocities due to the collected noise during high-frequency scanning, both the cell and the cargo trajectories were smoothed before data analysis by applying a moving average to each trajectory independently. To ensure that the major dynamics of the trajectories were captured also after smoothing, the length of the smoothing window was set equal to the decay time of the velocity correlation function of that trajectory, estimated as summarized in the following. Let the coordinates of the segmented object~(cell or cargo) in frame~$i$ be denoted by~$\vec{X}_{i} = \vec{x}_{i}+\boldsymbol{\sigma}_{i}$, where~$\vec{x}_{i}$ is the true position of the object and~$\boldsymbol{\sigma}_{i}$ denotes the imaging noise. The imaging errors in frames~$i$ and~$j$ are unbiased and uncorrelated 
\begin{equation}
\begin{gathered}
 \label{A:1}
   \langle \boldsymbol{\sigma}_{i} \rangle =0, \quad \langle \sigma_{i,\mu}\sigma_{j,\alpha} \rangle = s^2 \delta_{i,j} \delta_{\alpha,\mu},
\end{gathered}
\end{equation}
where~$s^2$ is the variance of the error from object tracking and~$\sigma_{i,\mu}$ denotes the~$\mu$\ts{th} Cartesian component of the noise in frame~$i$. Therefore, the~$i$\ts{th} measured velocity reads~$\vec{V}_{i} = (\vec{X}_{i+1} - \vec{X}_i)/\Delta t = \vec{v}_{i}+\boldsymbol{\eta}_{i}$, where~$\Delta t$ denotes the time step,~$\vec{v}_{i}$ is the true secant velocity and~$\boldsymbol{\eta}_{i}$ is the error of velocity~$\vec{V}_i$. The properties of the imaging noise~[Eq.~(\ref{A:1})] imply
\begin{subequations}
 \label{A:2}
\begin{align}
   \langle \boldsymbol{\eta}_{i} \rangle &=0 ,\\
   \langle \eta_{i,\mu}\eta_{j,\alpha} \rangle &= \frac{s^2}{\Delta t^2} \delta_{\mu,\alpha} \Big [2\delta_{i,j}-\delta_{i+1,j}-\delta_{i,j+1} \Big ] \! .
\end{align}
\end{subequations}
Using these definitions, the expectation value of the empirical velocity auto-correlation function for a trajectory with a total number of~$N$ frames can be written as follows: 
\begin{equation}
\label{eqn:evacf}
    \begin{aligned}
        \langle C_{\Delta} \rangle & =\frac{1}{N-\Delta}\sum_{i=1}^{N-\Delta} \langle \vec{V}_{i} \cdot \vec{V}_{i+\Delta} \rangle \\
                     & =\widetilde{C}_{\Delta}+\frac{2s^2}{\Delta t^2} \left (2\delta_{\Delta,0}-\delta_{\Delta,1} - \delta_{\Delta,-1} \right ).
    \end{aligned}
\end{equation}
where~$\widetilde{C}_{\Delta} = \sum_{i=1}^{N-\Delta} \vec{v}_{i} \cdot  \vec{v}_{i+\Delta} / (N-\Delta)$ for~$\Delta =0,1,2,...,N-1$ is the true velocity auto-correlation function.
Accordingly, the empirical auto-correlation function of the noisy secant velocities~$\vec{V}_i$ is an unbiased estimator of the correlation function of~$\vec{v}_i$ for~$\Delta  \ge 2$, since imaging noise does only affect the first two values~$\widetilde{C}_{\Delta = 0}$ and~$\widetilde{C}_{\Delta = 1}$. 
We fitted an exponentially decaying function to the first~$100$~points of the velocity correlation function for~$\Delta \ge 2$ and extrapolated for~$\Delta=0$ and~$\Delta=1$. 
The length of the smoothing window was chosen to be equal to the decay time~$\tau_c$ of the fitted exponential, proportional to~$e^{-t/\tau_c}$. 
For an example of the velocity correlation function and the exponential fit, both for the cell and the cargo, see Fig.~\ref{fig:S1} in the Supporting Material. 
Depending on the trajectory, the length of the smoothing window varies from $10 \, \mbox{s}$ to $30\, \mbox{s}$ for the cell and $5\, \mbox{s}$ to $10\, \mbox{s}$ for the cargo trajectories, respectively.

All regressions shown in the main text~[Figs.~\ref{fig:1}(f), \ref{fig:5}(b,c)] were performed by minimizing the reduced chi-square statistics~(mean squared weighted deviation).


\appendix

\newpage
\cleardoublepage
\renewcommand{\thefigure}{S\arabic{figure}}
\renewcommand{\thetable}{S\arabic{table}}
\setcounter{figure}{0}
\setcounter{table}{0}

\setcounter{page}{1}

\onecolumngrid
\normalsize
\nolinenumbers

\section{Description of movies}

\textbf{Movie 1:} An example of a cell carrying a spherical polystyrene particle with a radius of~$R=22.5 \, \mbox{\textmu m}$. The F-actin of the cell is labeled in red. The cell and the cargo trajectories are shown in red and blue, respectively.  \\ 
\textbf{Movie 2:} An example of a cell carrying a spherical polystyrene particle with a radius of~$R=14.5 \, \mbox{\textmu m}$. The F-actin of the cell is labeled in red. \\ 
\textbf{Movie 3:} An example of a cell carrying a spherical polystyrene particle with a radius of~$R=30.5 \, \mbox{\textmu m}$. The F-actin of the cell is labeled in red. \\ 
\textbf{Movie 4:} An example of a cell carrying a spherical polystyrene particle with a radius of~$R=50 \, \mbox{\textmu m}$. The F-actin of the cell is labeled in red. \\ 
\textbf{Movie 5:} An example of a cell carrying a spherical polystyrene particle with a radius of~$R=62 \, \mbox{\textmu m}$. The F-actin of the cell is labeled in red. \\ 
\textbf{Movie 6:} An example of a cell carrying a spherical polystyrene particle with a radius of~$R=73.5 \, \mbox{\textmu m}$. The F-actin of the cell is labeled in red. \\ 
\textbf{Movie 7:} A cell in a microfluidic channel carrying a spherical polystyrene particle with a radius of~$R=23\, \mbox{\textmu m}$. The flow direction is from bottom to top with a peak speed of 0.75~mm/s. The F-actin of the cell is labeled in red. \\ 
\textbf{Movie 8:} A cell in a microfluidic channel carrying a spherical polystyrene particle with a radius of~$R=23\, \mbox{\textmu m}$. The flow direction is from bottom to top with a peak speed of 0.75~mm/s. The F-actin of the cell is labeled in red. 

\section{Supplementary Figures}

\begin{figure*}[!ht]
\centering
\includegraphics[width=.8\textwidth]{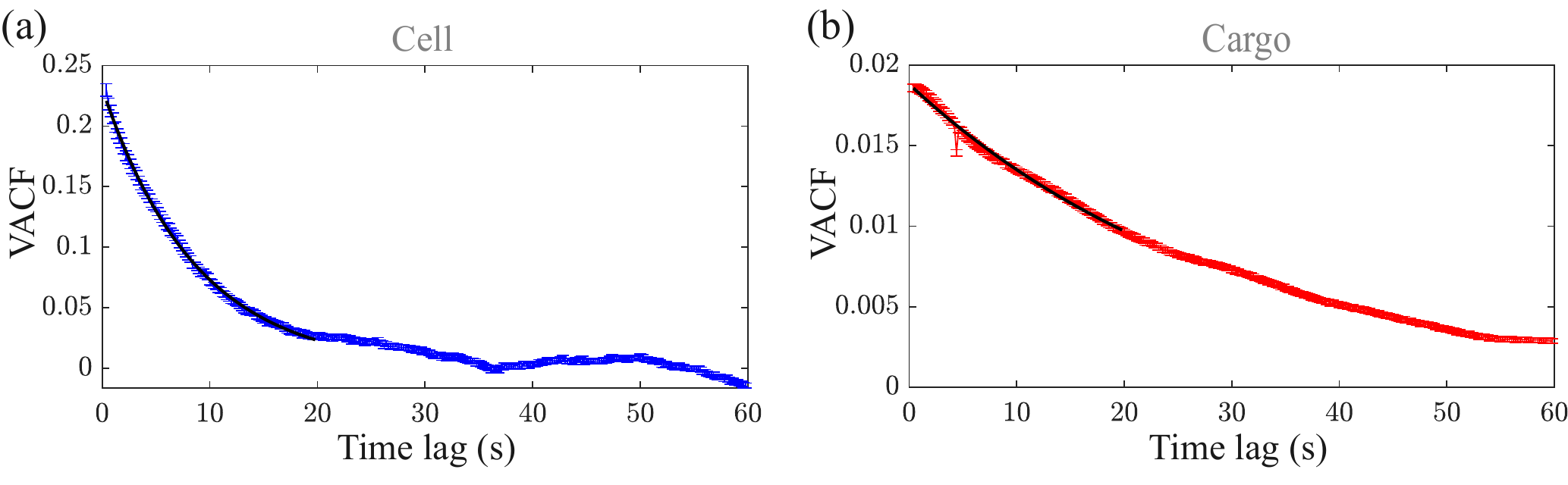}
\vspace{-0.2cm}
\caption{\fighead{Velocity autocorrelation function of the cell and the cargo} (a)~Velocity autocorrelation function~(VACF) of the cargo corresponding to the blue track in Fig.~\ref{fig:2}(a). (b)~VACF of the carrier cell that corresponds to the red track in Fig.~\ref{fig:2}(a). The solid back lines show the exponential fit to the VACF from~$T=0.4 \, \mbox{s}$ to~$T=20 \, \mbox{s}$. }
\label{fig:S1}
\end{figure*}

\begin{figure*}[!ht]
\centering
\includegraphics[width=.5\textwidth]{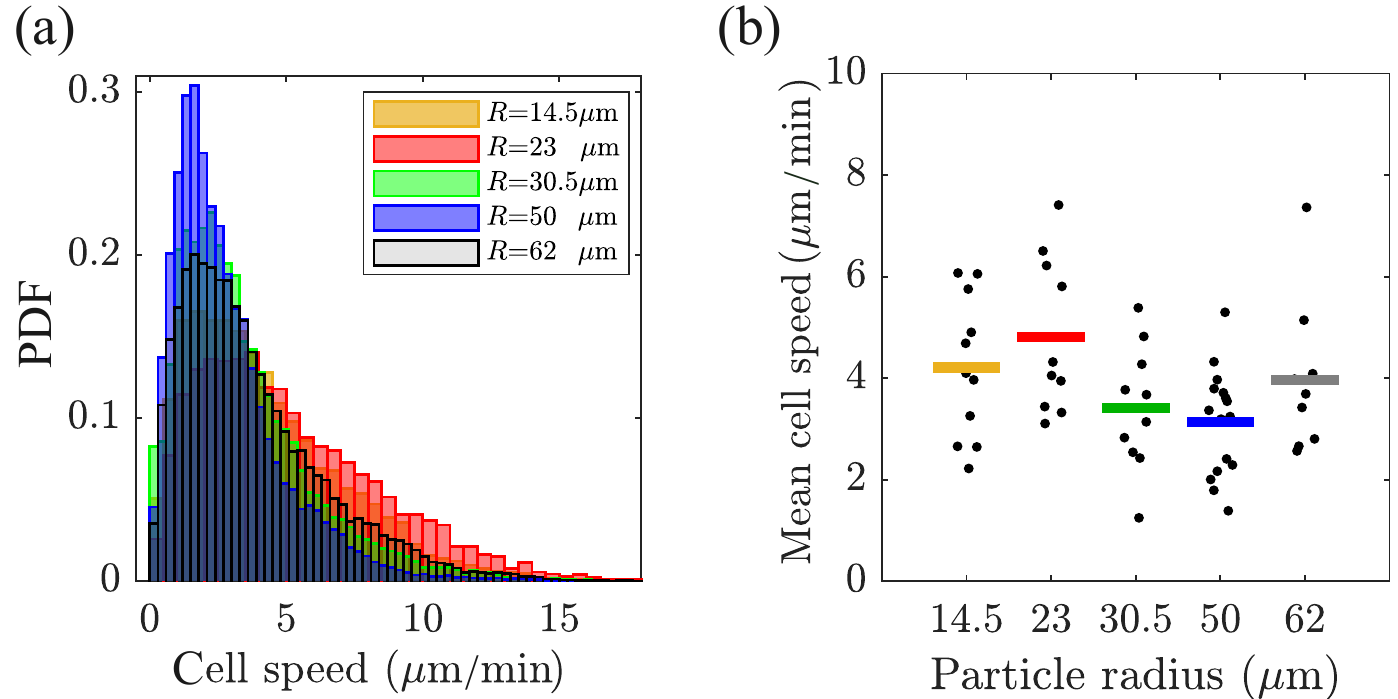}
\vspace{-0.2cm}
\caption{\fighead{Speed statistics of the carrier cell} (a)~The speed distribution of the carrier cells for each group of measured particle radii. (b)~The mean cell speed of each trajectory is depicted as black data points for the following particle radii:~$R=14.5$~\textmu m ($n=11$),~$R=23$~\textmu m ($n=10$),~$R=30.5$~\textmu m ($n=10$),~$R=50$~\textmu m ($n=16$) and,~$R=62$~\textmu m ($n=9$). The color-coded lines represent the averaged mean speeds of the carrier cell for each group of particles (color-code is identical to the colors in panel~a). }
\label{fig:S2}
\end{figure*}

\begin{figure*}[!ht]
\centering
\includegraphics[width=.6\textwidth]{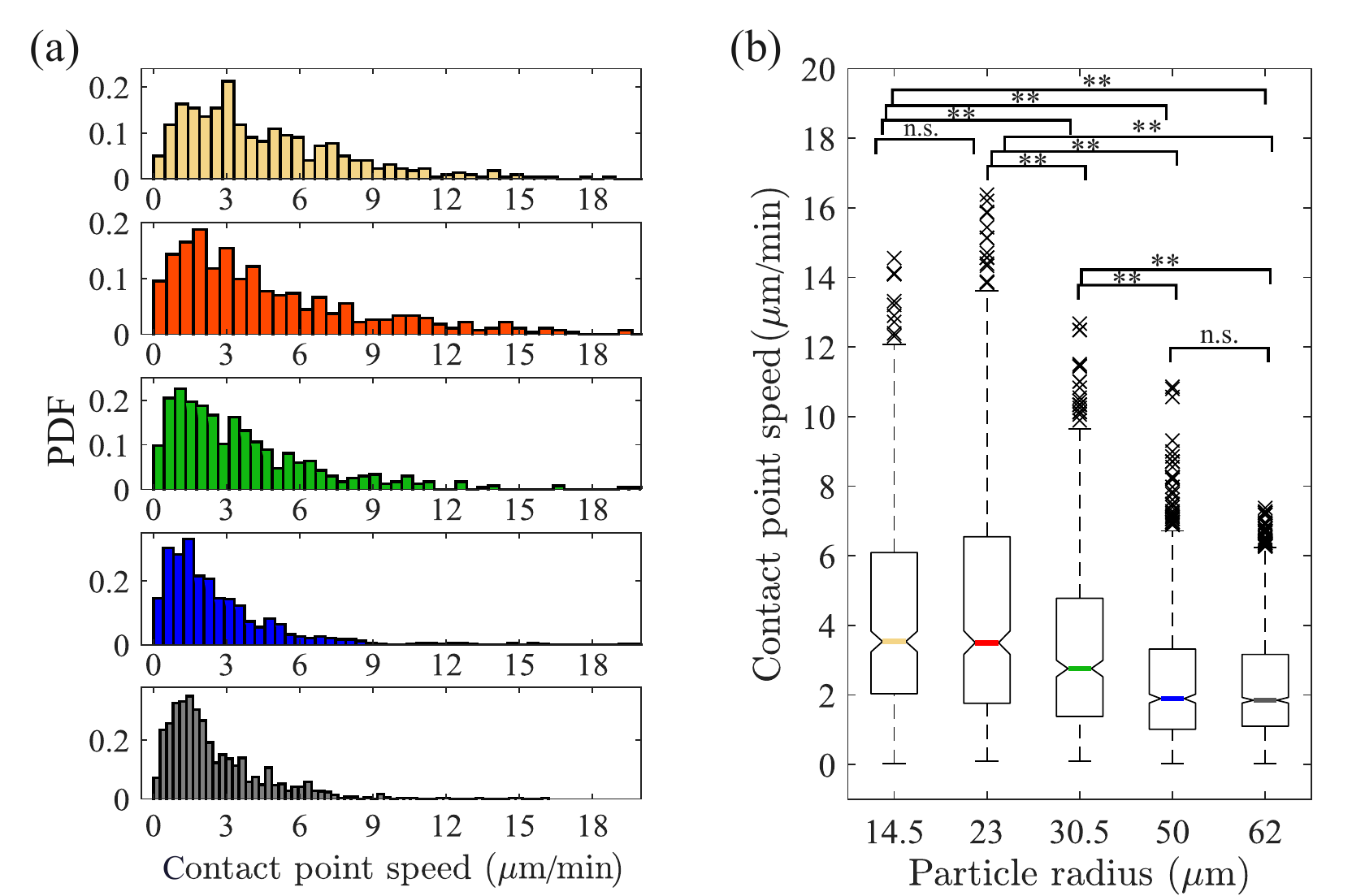}
\vspace{-0.2cm}
\caption{\fighead{Speed statistics of the contact point} (a)~Contact point speed distribution. From top to bottom, the speeds are calculated for contact points of particles with a radius of~$R=14.5$~\textmu m,~$R=23$~\textmu m,~$R=30.5$~\textmu m,~$R=50$~\textmu m and,~$R=62$~\textmu m. (b)~Decrease of the contact point speed as a function of particle radius. The color-coded lines show the median of the respective distributions. The upper and lower boundaries of each box mark the 75\ts{th} and 25\ts{th} percentiles, respectively~--~the distance between the top and bottom whiskers of each box thus shows the interquartile range. Brackets with two starts (\text{**}) indicate statistically significant differences (two-sample Kolmogorov-Smirnov test at a significance level of~$\alpha = 10^{-3}$); \textit{n.s.} stands for \textit{not significant}. }
\label{fig:S3}
\end{figure*}

\end{document}